**Practical language based on systems of definitions**

R. Nuriyev (renat.nuriyev@gmail.com)

**Abstract**

The article suggests a description of a system of tables with a set of special lists absorbing a semantics of data and reflects a fullness of data. It shows how their parallel processing can be constructed based on the descriptions. The approach also might be used for definition intermediate targets for data mining and unstructured data processing.

**Introduction**

The target of this article is to evaluate the practical usefulness of system of definitions. The following system called Dictionary Driven Reports (DDR) with a business data processing language shows how the system of named set of data can be used for application domain description, how much semantics it may carried even if its predicates are simple and how small additional info are needed to describe data fullness and to specify algorithms to process.

DDR considers office data as a system of tables that has to be verified, checked for the completeness and processed together.

Such description might be useful for data mining and an unstructured data processing[1]. A system of named tables is what holds main and easy to process information in economic, financial, scientific and engineering texts [2]. The value of tables can't be evaluated without knowledge of a system behind them. Probably reconstruction of the structure of the system is a one of fundamental goals of data mining.

**The first idea** considering here is that systems of tables are defined via four hierarchical named lists: list of names for the particular type of tables, hierarchical list of columns names, hierarchical list of names for rows and list of table attributes which carry context data related to the all data in table.

Type and name fully identify the table instance. One type table with multiple names we also call a **table series**.

**The second idea** is that a table processing is not an arbitrary – output tables have to be homomorphism images of input tables. If there is no "natural" homomorphism – then one table can't be got from others. Other word, the representation determines semantic of the tables.

**The third idea** is that cells of input tables that mapping to one output cell might be processed independently in parallel.

So it is an example of parallel programming without using a word "parallel" or "synchronization".

For the next table reflecting office expenses we have seven lists

DEPARTMENT={IT, HR},
QUARTERS={1,2,3,4},
YEARS={2007,2008},
PROJECTS={pr1,pr2},
EXPENSE={office, trip}.
PERSONEL={p10,p11,p21,p22, p23}
INDICATOR={expenses, personnel, amount, max personnel, max amount}.
ADDRESS = {zip code, city, street}.

Then expenses table is:

names=DEPARTMENT×QUARTER×YEAR
columns=INDICATOR
rows=PROJECTS//EXPENSE//PERSONEL
table attributes = ADDRESS× DEPARTMENT × QUARTER × YEAR.

Next is an example of such table:

type: EXPENSE REPORT
name:-
table attribute:

| Address | Depart | quarter | year |
|---------|--------|---------|------|
| 08550   | IT     | 3       | 2007 |

Row/columns:

| projects | amount |
|----------|--------|
| pr1      | $12    |
| office   | $8     |
| p11      | $3     |
| p01      | $5     |
| trip     | $2     |
| p11      | $2     |
| pr2      | $7     |
| office   | $4     |
| p21      | $4     |
| trip     | $3     |
| p23      | $3     |

For a table definition it can be used expressions with regular list operations and some additional ones. In the example above table attributes defined as a Cartesian product of four lists

ADDRESS╳DEPARTMENT╳QUARTER╳YEAR, list of rows is constructed with a special operation '//' to get a plane list from hieratical lists PROJECTS//EXPENSES//PERSONAL.

If we have a table and its lists description we may check if personal IDs are correct, departments and date are OK. We may check if we got all reports from each department and for current quarter and year. Also we may check if its processing is correct using additional info about **Suppes - Zinnes measurement scales** – set of operations which make sense to do.

For processing column "amount" it make sense summation, multiplication to constant (to convert one currency to another), but not allowed multiply amount $A to amount $B.

It make sense a square feet but does not make sense a square dollars or kilograms. But the operation of division amount to amount is made sense.

Here the DEPARTMENT list has a simplest scale – individual, for which only operations $=, \neq$ and counting elements are allowed. The QUARTER list has a comparison scale: in addition to previous operation it has a $<, >$ operations. The column "amount" has a scale with all operations above and also multiplication to constants, dividing to number from personnel column, summation, max, min and so on.

For a hierarchical list of rows we allow summation operations only for all items inside one level hierarchy, but do not allow mix different levels or use part of them. Such sort of information has a great deal of semantic and helps to build **intellectual user interface**.

The transposition of columns and rows does not change the information in tables and processing algorithms.

For **data mining** algorithms it gives targets for further search: how tables are constructed, what lists have to be found, what is the (operational) semantics of data, how to check a completeness of data (if tables or some part of table are missing) and their correctness.

For a **data verification and data completeness** the description allow to check if all reports from each department and current quarter are received (list of departments DEPARTMENT and list of quarters QUARTER) and to avoid misspelling user has only to select. The list of absent reports can be generated as well. Also a report for the same quarter of last year can be defined. For additional data checking it can be used a test that numbers are not differ for more than 15% from numbers of the previous report.

The important feature of the LDR is that data has a human representation – two dimensional tables with 4000 years of experience and polishing.

Further we will use more specific type of lists and call them **dictionaries**.

Core of the DDR system are notions of local universes (sets of elements for dictionaries and tables), dictionaries (special type of named lists with elements from some universe) and tables. Examples of local universes are a list of employee names, list of cities in USA. Dictionaries are subsets of these universes. Example of such subset is a list of IT department employee's names. For minimizing input and avoiding misspelling subsets are identified with universe's name and set of indexes of its elements in this universe.

There are "a priory" given local universes
- NaturN is a set of nutural numbers,

- RealR is a set of real numbers,

- CharC is a set of strings of characters

- Intervals is set of pairs of numbers or strings.

More complex types of universes are vectors and graphs.

Table definition dictionaries might be defined as results of dictionary operations.

**Def. 1**. . A **dictionar**y has a **name** (string of characters) and consists of two components:

-  name of local universe,

-  list of indexes for elements of this universe accompanied with genesis - sets of dictionary names (and hence a universe name) to which the element belongs.

DDR applications have a set of operations operS - a 3D vector (function name, list of local universes for arguments, list of local universes for results).

The operation is **applicable** if input cell dictionaries contain dictionaries in operation arguments and output cell contains output dictionaries.

DDR Design Studio consists of several subsystems. One of them is a dictionaries and tables definitions subsystem. It supports inputting named dictionaries, creating new dictionaries with operations and defining tables with four dictionaries – dictionary of names, dictionary of table attributes, dictionary of column names and dictionary of row names. The subsystem checks if all of them are filled up and if not – force to do it. If all dictionaries are filled up, it checks if all tables with possible names are filled. If not – gives a list of absent tables.

Dictionaries might be defined as result of operations and its combinations above anther dictionaries.

**Def. 2. Dictionary operations** are set theory operations ∪, ∩, ×, +, - for pair arguments, lexicographical and numerical orderings and several additional operations:

- **creating hierarchy of named** dictionaries**:** expression <dictionary A>(<dictionary B>,…,<dictionary C>); it defines a set of (empty) dictionaries with names A(b,…, c) for each combinations of elements of B,…, C; B,…,C are dictionaries or dictionary expressions with operations,

- **transformation serial dictionaries to another serial** dictionary: expression <operation> <hierarchical dictionaries>| <dictionary1>…| <dictionaryN>; here <operation> is one of symbols ∪, ∩, ×, +, - , <hierarchical dictionary> is a hierarchy of named dictionaries, dictionary1…, dictionaryN are one of argument in it expression; operation produces a several dictionaries with names O(∪ A(B, $c_1$)) ,…,O(∪ A(B, $c_m$)); each of them is a joint of all dictionaries of A($b_i$, $c_p$) for each $b_i$∈B and fixed $c_p$,

- **expansion of hierarchical dictionary:** expression <hierarchical dictionary E>(<dictionary A>,…, <dictionary B>)//<dictionary C>//… define a dictionary of elements: $c_1$, followed with <joint of all dictionaries with any element of A,…, any element of B and fixed element $c_1$ from C>, then next element $c_2$∈C again follower with <joint of all dictionaries with any A,…, any element of B and fixed element $c_2$ from C>, the again c3,…..

Each elements of any dictionary keeps track of its origin via a value of special attribute called **genesis**. For example, element w of A∪B is **genesis** has A, B if w belongs to A and B; if it belongs only to B, it is B only. These attribute is used for processing as well.

**Examples of operations**. Denote a PERSONAL list via P, a DEPARTMENT={IT, HR} via D, BRANCH={USA, Can, EMU} via B. The expression P(D, B) defines names for 6 personal lists: P(IT, USA), P(IT, Can), P(IT, EMU), P(HR, USA), P (HR, Can), P(HR, EMU).

The expression P (D, B)|D defines three dictionaries - one for each branches without dividing by departments:

Dictionary1: name = P(USA) , elements={(Person1, **generics** USA∈B, IT∈D(USA)), (Person2, **genesis** USA, IT∈D(USA)), (Person4, **genesis** USA∈B, HR∈D(USA))};

Dictionary 2: name = P(Can), elements={ (Person3, **genesis** Can∈B, IT∈D(Can))};

Dictionary 3: name = P(EMU), elements={( Person5, **genesis** EMU∈B, HP∈D(EMU))}.

The expression P(D, B)//D gives a plain dictionary:

IT, **genesis** D(B)
Person1, **genesis** USA∈B, IT∈D(USA)
Person2, **genesis** USA ∈B, IT∈D(USA)
Person3, **genesis** Can∈B, IT∈D(Can)
HR, **genesis** D(B)
Person4, **genesis** USA∈B,HR∈D(USA)
Person5, **genesis** EMU∈B, HR∈D(EMU)

It consists of sections starting with department names and followed by a list of people from departments of each branches.

**Table series** are defined by types and four dictionaries for defining its structure and names:
- dictionary of names,
- dictionary of table attributes,
- dictionary of columns,
- dictionary of rows.

These dictionaries might be defined with expressions as well, but outside of tables.

Elements of table series dictionaries are **selected in parallel**, meaning if a dictionary name is in two places of expressions, then the same element will represent this list.
Each element of name dictionary determines a single table with rows and columns.

**Example 1.** Let T be a table series:

name = A
table attributes = U
rows = R(A, B)//B
columns = C(A).

Then a single table with name *a* has columns from C(*a*) and rows from list R(*a*, B)//B. Generally, for different names the same type of table may have different columns and rows.

An elementary table object is a cell. It is identified by an element of name list, values of table attributes, element of row list and element of column list.

Some table series are a little bit unusual:
name: -
table attributes: -
rows : R
columns : C(R).

Here each row has its own set of columns.

More studies are needed to determine a more precisely meaning and sense of table series definitions.

As it was mentioned before, together with table processing applications it makes sense to have a collection of operations applicable to columns and rows from of the same or different table series. This set of operation defines all possible ways for table series processing and it may be considered as an operational semantics of data and possible ways for future transformations. It also works for semantic errors detection in a future application evolution.

**2. Table processing**

**Def. 3. Table processing** is a mapping several table series, called **input tables**, to one table series, called **result table**. The mapping is based on dictionaries: the processing is allowed only for cells with consistent genesis when each dictionary from result tables cell's genesis is in genesis of input cells.

**Remark 1**. It make sense to have lists in description of output tables which are not in descriptions of these input tables and to keep its cells unchanged after this processing - they might be used for later processing or input.

**Example 2.** Subject area for this example is determined by dictionaries.
Company structure is determined via the following dictionaries:
BRANCHES = {br1,br2,br3}
and
DEP(BRANCHES) contains three dictionaries:

DEP(br1)={d11,d12,d13},
DEP(br2) ={d21,d22},
DEP (br3)={d31}.

Projects are determined with TYPES ={ r, s, h} (r-research, s-software, h-hardware), and departments DEP.

PROJECTS(DEP(BRANCH), TYPES)={PROJECTS(d11, r)={p1,p2}, PROJECTS(d12, r)={r3}, PROJECTS (d13, r)={r4},…, PROJECTS(d13,h)={r3,r4}} is lists of all projects for each BRANCH, DEPartment and project TYPES.

Input report tables are quarterly department reports:

type:  EXPENSE-REPORT
names : -
table attributes: DEP, BRANCH, QUARTER;
columns: INDICATORS;
rows: PROJECTS(DEP(BRANCH), TYPES));

where INDICATORS ={ expense, personal}.

The following is one of the tables of series EXPENSE-REPORT:

Table attributes

| Dep | Branch | Quart |
|-----|--------|-------|
| d21 | br2    | 3     |

| projects | expenses | personal |
|----------|----------|----------|
| p21 | 3 | 2 |
| p22 | 4 | 1 |
| P23 | 2 | 1 |
| P24 | 2 | 1 |

Genesis for a cell (p22, expenses) of this table is "PROJECTS p22, DEP d21, BRANCH br2, TYPE r, QUART 3", value of the cell is 4.

The instance of the output report table:

> type: **Expense-by-branches**
> name**:-**
> table attributes:-
> column: INDICATORS,
> rows: BRANCHES

is shown bellow:

| branches | expenses | personel |
|---|---|---|
| br1 | 11 | 16 |
| br2 | 8 | 10 |
| br3 | 20 | 17 |

For cell (br1, expenses) of this table genesis is: "INDICATORS expenses, BRANCHES br2".

Collection of attributes for first table genesis includes the attributes of the second one. So there is a homomorphism of EXPENSE-BY-BRANCHES cells (department, branches, projects, expenses, personal) to EXPENSE-REPORT cells (expenses, personal, branches) and values of these cells can be process together.

For complete determination of the processing we need to add that to cells of columns 'expense' and 'personal' are applicable accumulative functions:

Expense-by-branches(expenses, branches) = **accum**(EXPENSE-REPORT(expense, branches),

Expense-by-branches(personal, branches) = **accum**(EXPENSE-REPORT(personal, branches).

Then all cell's values of homomorphism images are summarized to cells (expenses, branches) and (personal, branches) and we will get a final table of expenses and personal used in projects for each branches.

Adding "efficiency" column to table EXPENSE-BY-BRANCHES allows to

obtain output table of type EXPENSE-EFFICIENCY with a column dictionary ADD-IND = INDICATORS ∪ {efficiency}:

type: EXPENSE-EFFICIENCY
names: -
table attribute: -
columns: ADD-IND
rows: BRANCHES,

The computing formula is:
EXPENSE-EFFICIENCE(efficiency)=
EXPENSE-BY-BRANCHES (expenses) / EXPENSE-BY-BRANCHES(personal).

**Remark 2**. Table series approach gives a formal base for defining important and illusive thing such as **full definition of subject matter**.

**Remark 3**. Another important fact is that table series definition in large part is defined the **algorithm of its processing**. So program of it processing became very small. Less debugging is needed.

**Remark 4.** For analysis **unstructured text mining**, reconstruction a system of connected table may be a **strategic target**, determining if and when we got all available info. Reconstructing a subject area – full set of tables without holes – is much more important than collecting pieces of unknown size of information we got in one separate table.
This also determines an **intermediate targets** for mining: lists for table names, columns and rows, tables with common lists.

**Remark 5.** Many data have a temporary importance, and it is often hard and expensive to carry ballast data. Considering a business data as a flow of permanently coming and disappearing data (not all of it has to be accumulated) this description will help to build more adequate systems than one based on database concept.

Such **application** is a set of local table rules (LTR) and reserved (system) table LRRE of local rule names ready for execution. Rules have names and two collections: list of input tables and list of output tables. They are form a string in reserved table LTP-D with 3 columns: "LTR-name" for local table rule name, "LIT" for list of input tables and "LOT" for list of output tables. The final reserved table is LTP-In with three columns: "LTP-P name", "Initializing type" and "Initializing name".

The DDR Monitor starts when LRRE table is not empty. The LRRE table is updated if some ordinary application tables are updated or have been input by user or came from outside networks. Rule with empty initialized list also starts the application but never is executed again because input tables never updated - there are no ones. When started, the application never stops, it may only wait for updating some of input tables. After inputting or editing or creating table the Monitor checks the LTP-D table and adds these rows to LRRP. Then it executes rules from its rows with updated initialized tables.

If for some reason computed was restarted, the service again starts with LRRE. Because of any local rule does not change any application table - they might be updated only with the service - after restarting the computer, any unfinished steps will be repeated without any data corruption (as MSMQ does).

The DDR itself might functioning and growing the same way. For example, to add a distributed functionality it is enough to have a (system) rule for sending/receiving data to/from another nodes. To support securities, the only (system) rule to be initialized is a rule for checking authorization and outputting a LRRE.

This project might be good for student themes. It does not need any DB, complex OS and programming languages or any other expensive software. Working prototype might be finished in 3-5 months in LINUX and C++ or JAVA.